\documentclass[aps,superscriptaddress,twocolumn,twoside,floatfix,pra,nofootinbib,a4paper]{revtex4-2}

\pdfoutput=1

\usepackage{times}
\usepackage{amsfonts}
\usepackage{amsmath}
\usepackage{amssymb}
\usepackage{amsthm}
\usepackage{multirow}
\usepackage[normalem]{ulem}
\usepackage[T1]{fontenc}
\usepackage{float}
\usepackage{mathtools}
\usepackage{bm}
\usepackage[UKenglish,cleanlook]{isodate}   
\usepackage{mathrsfs}
\usepackage{dsfont}

\usepackage{physics}

\usepackage{xcolor}
\definecolor{carmine}{RGB}{150,0,24}

\usepackage{natbib}
\usepackage[colorlinks=true,linkcolor=blue,citecolor=magenta,urlcolor=blue]{hyperref}

\newcommand{\bracket}[3]{\langle#1|#2|#3\rangle}

\newtheorem{result}{Result}
\usepackage{caption}
\usepackage{graphicx}
\usepackage{tikz-cd}

\begin{document}


\title{Resource-efficient high-dimensional entanglement detection via symmetric projections}

\author{Simon Morelli}
\affiliation{BCAM - Basque Center for Applied Mathematics, Mazarredo 14, 48009 Bilbao, Spain}

\author{Marcus Huber}
\affiliation{Atominstitut, Technische Universit\"at Wien, 1020 Vienna, Austria}

\author{Armin Tavakoli}
\affiliation{Physics Department, Lund University, Box 118, 22100 Lund, Sweden}

\date{\today}

\begin{abstract}
We introduce two families of criteria for detecting and quantifying the entanglement of a bipartite quantum state of arbitrary local dimension. The first is based on measurements in mutually unbiased bases and the second is based on equiangular measurements. Both criteria give a qualitative result in terms of the state's entanglement dimension and a quantitative result in terms of its fidelity with the maximally entangled state. The criteria are universally applicable since no assumptions on the state are required. Moreover, the experimenter can control the trade-off between resource-efficiency and noise-tolerance by selecting the number of measurements performed. For paradigmatic noise models, we show that only a small number of measurements are necessary to achieve nearly-optimal detection in any dimension. The number of global product projections scales only linearly in the local dimension, thus paving the way for detection and quantification of very high-dimensional entanglement. 
\end{abstract}

\maketitle

\textit{Introduction.---}
Entanglement is a paradigmatic resource in quantum information science. It is essential for applications in communication \cite{DenseCode, Teleport, Buhrman2010},  cryptography \cite{Scarani2009, Xu2020, Pirandola2020},  sensing \cite{Degen2017, Giovannetti2011} and device-independent information processing \cite{Pironio2010, Acin2007}. It is also crucial for fundamental tests, for example in nonlocality \cite{Brunner2014, NetworkReview} and for the nature of gravity \cite{Bose2017}. Therefore, entanglement has received massive research attention \cite{Plenio2007, Horodecki2009, Guhne2009, Friis2019}.

An important frontier is the entanglement between two high-dimensional systems. It is well-known that entanglement typically becomes much more robust to noise as the dimension increases. This enables stronger tests of steering \cite{Marciniak2015, Srivastav2022} and even advantages in quantum nonlocality \cite{Collins2002, Dada2011}, leading to device-independence for high-dimensional systems \cite{Salavrakos2017, Tavakoli2021b, Sarkar2021}. In quantum key distribution, higher-dimensional entanglement can lead to higher key-rates \cite{Sheridan2010, araujo2023, Cozzolino2019, Bouchard2018, Bulla2023} and to higher tolerance of errors for the security \cite{Cerf2002}. In entanglement-assisted quantum communication, it can boost the advantages of qubit messages \cite{Tavakoli2021, Pauwels2022}, increase the capacity of a quantum channel \cite{Hu2018} and enhance the noise-tolerance of teleportation \cite{Luo2019}. Naturally, high-dimensional entanglement has been the focus of many optics experiments \cite{Erhard2018, Erhard2020}. Realisations have been reported for instance in transverse spatial modes  \cite{Krenn2014, Bavaresco2018, Fontaine2019, HerreraValencia2020}, in path \cite{Wang2018, Hu2020}, in time-bins \cite{Martin2017, Ecker2019} and in frequency modes  \cite{Kues2017, Reimer2016, ponce2022unlocking}. Beyond optics, higher-dimensional entanglement has been developed in e.g.~trapped ion quantum computers \cite{Ringbauer2022} and superconducting circuits \cite{Lierta2022}.

In view of all this progress, a central challenge is to develop methods for detecting and characterising entanglement. A common approach is to perform state tomography and then apply a suitable entanglement criterion to the reconstructed density matrix. However, this is typically only viable for low-dimensional systems; partly because entanglement detection is difficult even with the density matrix in hand \cite{Gurvits2004, Gharibian2010}, but mainly because of the rapidly increasing resource cost. For a bipartite system of local dimension $d$, tomography requires measurements in $(d+1)^2$ global product bases. Since for many optical platforms, especially when $d$ is large, it is a challenge to simultaneously resolve all $d$ possible local outcomes, it is often more relevant to perform $d^2(d+1)^2$ local filter settings, i.e.~to make global product projections separately onto each component of the basis. This often considerably simplifies experimental requirements. A different approach is to detect entanglement via the fidelity of the unknown state with the maximally entangled state \cite{Bourennane2004}. This is informative because  often the most useful entanglement is close to the maximally entangled state, which considerably narrows the otherwise much larger set of entangled states  \cite{Weilenmann2020}. By performing suitable global product projections, one can deduce the fidelity and thereby quantify the entanglement. Moreover, from the fidelity one also obtains a lower bound on an important  qualitative property of the state, namely the number of entangled degrees of freedom (dimension) needed to prepare the state. This is known both as the entanglement dimension and Schmidt number \cite{Terhal2000}. While the fidelity can be deduced using much fewer measurements than tomography, it is still resource-intensive for larger dimensions, requiring $d(d+1)$  global product projections (filters) if the total count rate is known.
It  motivates the need for more efficient approaches to fidelity-based entanglement detection.

Notably, the practical difficulties of certifying high-dimensional entanglement have sometimes motivated  conveninent additional assumptions to simplify the problem. Here, we will make no such assumptions. Thus, we develop an approach that is valid independently of the precise physical modelling of the state. In this setting, we introduce two  practically useful classes of entanglement detection criteria. They both provide lower bounds on the fidelity with the maximally entangled state and a lower bound on the Schmidt number. One criterion is based on mutually unbiased bases (MUBs). The other criterion is based on equiangular measurements (EAMs), among which the most well-known example is the symmetric informationally complete measurement (SIC-POVM) \cite{Renes2004, Zauner2011}. Both these classes of measurements are broadly relevant in quantum information science and they are frequently studied in both theory \cite{Durt2010, Fuchs2017} and experiment \cite{Mafu2013, HerreraValencia2020, Srivastav2022, Medendorp2011, Bent2015, Smania2020, Huang2021, Sticker2022}.

In addition to their universality, both our criteria have three important practical features. Firstly, they are versatile. The experimenter can freely choose how many global product projections are implemented, thus tuning the trade-off between using few measurements and tolerating large amounts of noise \cite{Liu2022}. Secondly, they require very few projections. The experimenter does not need to measure global product bases. They need only to estimate the total count rate and measure the much smaller subset of global filter projections corresponding to identical outcomes. Thus, no data needs to be collected for non-identical outcomes. For example in the case of MUBs, and similarly for EAMs, this reduces the scaling of the number of projections to being only linear in $d$. This not only greatly improves on the above fidelity discussion but also on comparable criteria \cite{Erker2017, Bavaresco2018}, thus making viable tests of very high-dimensional entanglement. Thirdly, for standard noise models, only a small number of projections are necessary to obtain nearly optimal noise-tolerance. This means that the large savings in resource cost come at a much smaller cost in accuracy.

\textit{Schmidt numbers and entanglement fidelity.---} The number of entangled degrees of freedom in a pure bipartite state $\ket{\psi}_{AB}$ of local dimension $d$ is given by its Schmidt rank. That is the number of terms, $r(\psi)$, appearing in the Schmidt decomposition $ \ket{\psi}=\sum_{i=1}^{r(\psi)} \lambda_i \ket{\alpha_i,\beta_i}$, where $\{\ket{\alpha_i}\}_i$ and $\{\ket{\beta_i}\}_i$ are, respectively, orthonormal states and $\{\lambda_i\}_i$ satisfy $\lambda_i> 0$ and $\sum_i \lambda_i^2=1$. The Schmidt rank is an integer delimited by $1\leq r\leq d$, with $r=1$ ($r=d$) meaning that the state is product (fully entangled) and $1<r<d$ meaning that  entanglement is present but confined to a smaller subspace.  For mixed states, $\rho_{AB}$, the Schmidt rank generalises to the Schmidt number  \cite{Terhal2000}. The Schmidt number, $k(\rho_{AB})$, is the largest Schmidt rank of all the pure states $\{\ket{\psi_i}\}$ appearing in a given convex decomposition of $\rho_{AB}$, minimised over all possible decompositions. Thus,
\begin{align}\nonumber
k(\rho_{AB})\equiv \min_{\{q_i\},\{\psi_i\}}  \Big\{r_\text{max}:& \quad  \rho_{AB}=\sum_i q_i \ketbra{\psi_i}\\\label{SN}
&\text{and} \quad r_\text{max}=\max_i r(\psi_i)\Big\}.
\end{align}

While the Schmidt number provides a qualitative benchmark for the extent to which a high-dimensional state is entangled, it does not mean that the entanglement is useful. For example, the state $\ket{\psi}=\sqrt{1-\epsilon}\ket{00}+\sqrt{\frac{\epsilon}{d-1}}\sum_{i=1}^{d-1}\ket{ii}$ has  maximal Schmidt rank ($r=d$) for any  $0<\epsilon<1$ but in the limit $\epsilon\rightarrow 0$ it is arbitrarily close to the product state  $\ket{00}$ ($r=1$). Therefore, we also quantitatively study the entanglement, through its fidelity with the maximally entangled state,
\begin{equation}
F(\rho_{AB})\equiv  \max_{U_A} \bracket{\phi^+_d}{U_A\otimes\openone_B \rho_{AB} U_A^\dagger \otimes \openone_B }{\phi^+_d},
\end{equation}
where $U$ is a unitary operator and $\ket{\phi^+_d}=\frac{1}{\sqrt{d}}\sum_{i=0}^{d-1}\ket{ii}$. We will refer to $F(\rho_{AB})$ simply as the entanglement fidelity. Moreover, the entanglement fidelity implies a simple lower bound, $F(\rho_{AB})\leq  \frac{k(\rho_{AB})}{d}$, on the Schmidt number \cite{Terhal2000}.

\textit{Entanglement criterion via MUBs.---} A pair of bases are called mutually unibased if the modulus overlap between any two of their elements is constant. Similarly, a set of $m$ bases $\{\ket{e_a^z}\}$, indexed by $z=1,\ldots,m$ with basis element $a=0,\ldots,d-1$, are called MUBs if the unbiased property holds between every pair. That is, MUBs satisfy $\left|\braket{e_a^z}{e_{a'}^{z'}}\right|^2=\frac{1}{d}$ for any $z\neq z'$. For any $d$, at least  $m=3$ and at most $m=d+1$ MUBs exist. Saturation of the upper bound implies a tomographically complete set and it is known to be reachable in all dimensions that are powers of prime numbers \cite{Wootters1989}.

Towards detecting entanglement, consider that we perform global product measurements of the projectors comprising $m$ MUBs. Specifically, we measure $\ketbra{e_a^z}{e_a^z}\otimes \ketbra{e_a^{z*}}{e_a^{z*}}$, where $\ket{\psi^*}$ denotes the complex conjugate of $\ket{\psi}$. The set $\{\ket{e_a^z}\}$  is only assumed to satisfy the MUB-property. As our witness of entanglement, we use the sum-total of the probabilities that the local outcomes are identical, i.e.~
\begin{equation}\label{Wmub}
\mathcal{S}_{m,d}(\rho_{AB})=\sum_{z=1}^m \sum_{a=0}^{d-1} \bracket{e_a^z,e_a^{z*}}{\rho_{AB}}{e_a^z,e_a^{z*}}.
\end{equation} 
We choose this quantity for three reasons. Firstly, for any selected set of $m$ MUBs it is invariant under permutations of the basis label and the outcome label respectively.  Secondly, it is particularly well-suited for the most relevant entangled state, namely $\ket{\phi^+_d}$. Since this state is invariant under any local unitaries of the form  $U\otimes U^*$, it follows that perfect correlations must be observed in every product MUB. This leads to the algebraically maximal value $\mathcal{S}_{m,d}(\phi^+_d)=m$. Thirdly, $\mathcal{S}_{m,d}$ can be measured in the lab using few global filter projections (identical outcomes), as compared to measuring the full global product bases.

We now present our first criterion, showing that $\mathcal{S}_{m,d}$ can be used both to detect the Schmidt number of $\rho_{AB}$ and to bound its entanglement fidelity.
\begin{result}[MUBs]\label{result1}
For any bipartite state $\rho_{AB}$ of equal local dimension with Schmidt number at most $k$ it holds that
\begin{equation}\label{MUBentcriterion}
\mathcal{S}_{m,d}(\rho_{AB}) \leq 1+\frac{(m-1)k}{d}.
\end{equation}
Moreover, any observed value $\mathcal{S}_{m,d}$ implies the entanglement fidelity bound
\begin{equation}\label{MUBfidcriterion}
F(\rho_{AB})\geq \frac{\mathcal{S}_{m,d}-1}{m-1}.
\end{equation}
\end{result}
The proof is fully analytical and given in Supplementary Material. The special case corresponding to Eq.~\eqref{MUBentcriterion} and separable states ($k=1$) was obtained by a different proof method in \cite{Spengler2012}. Already $m=2$ MUBs are sufficient to detect the largest possible Schmidt number. However, by using more MUBs the gap between $\mathcal{S}_{m,d}(\phi^+_d)$ and the bound \eqref{MUBentcriterion} grows, indicating improved noise-robustness of entanglement detection. We return to the noise analysis later. Complementarily to our case, lower bounds on $\mathcal{S}_{m,d}$ for small $d$ were explored in \cite{Bae2019} and a modification of $\mathcal{S}_{m,d}$ can detect bound entanglement \cite{Bae2022}.

\textit{Entanglement criterion via EAMs.---} A set of $n$ pure states $\{\ket{\psi_a}\}_{a=1}^n$ of dimension $d$ is called equiangular if the modulus overlap between any pair of distinct states is constant, i.e.$\left|\braket{\psi_a}{\psi_{a'}}\right|^2=t_{n,d}$ for every $a\neq a'$. The constant cannot take a value smaller than $t_{n,d}=\frac{n-d}{d(n-1)}$ \cite{Welch1974}. The set forms a so-called equiangular tight frame if and only if this lower bound is saturated, meaning in particular that the subnormalised projectors $\{\frac{d}{n}\ketbra{\psi_a}\}$ form an equiangular quantum measurement. Considerable work has been directed at deciding the existence of EAMs (see e.g.~\cite{Lemmens1973, Fickus2015, SUSTIK2007}). In particular, when $n=d$ they reduce to an orthonormal basis. When $n=d+1$, an EAM is obtained from removing one row from the $(d+1)$-dimensional Fourier matrix and renormalising the columns. When $n=d^2$, EAMs are equivalent to SIC-POVMs. The latters are known to exist in every dimension up to at least $d=151$ \cite{Fuchs2017} and they are particularly interesting because the size of an EAM is delimited by $d\leq n\leq d^2$.

We now present an entanglement witness based on  local measurement of an EAM.  The main idea parallels that of Result~\ref{result1}. We consider global product projections $\ketbra{\psi_a}{\psi_a}\otimes \ketbra{\psi_a^*}{\psi_a^*}$ where $\{\ket{\psi_a}\}_{a=1}^n$ can correspond to any EAM. In analogy with \eqref{Wmub}, we consider the sum-total of the probabilities that the local outcomes are identical. Up to a conveninent constant, this is given by
\begin{equation}\label{Weam}
\mathcal{R}_{n,d}\equiv \frac{d(n-1)}{n(d-1)} \sum_{a=1}^{n} \bracket{\psi_a,\psi_a^*}{\rho_{AB}}{\psi_a,\psi_a^*}.
\end{equation} 
Again, for a given choice of EAM, this quantity is invariant under permutations of the outcome label and the maximal value (when $n>d$) is obtained from the maximally entangled state. One has
$\mathcal{R}_{n,d}(\phi^+_d)=\frac{n-1}{d-1}$. We now present our second main result; an entanglement criterion based on $\mathcal{R}_{n,d}$.

\begin{result}[EAM]\label{result2}
For any bipartite state  $\rho_{AB}$ of equal local dimension  with Schmidt number at most $k$ it holds that 
\begin{equation}\label{EAMentcriterion}
\mathcal{R}_{n,d}(\rho_{AB})\leq 1+k\frac{n-d}{d(d-1)}.
\end{equation}
Moreover, any observed value $\mathcal{R}_{n,d}$ implies the entanglement fidelity bound 
\begin{equation}\label{EAMfidcriterion}
F(\rho_{AB})\geq \frac{(d-1)(\mathcal{R}_{n,d}-1)}{n-d}.
\end{equation}
\end{result}
The proof is given in Supplementary Material and it is based on ideas that closely parallel those used in the proof of Result~\ref{result1}. It is interesting to note that already the smallest non-trivial EAM, namely $n=d+1$, is sufficient to detect a maximal Schmidt number. However, in analogy with the MUB-case, using more projections (larger $n$)  leads to a larger gap between $\mathcal{R}_{n,d}(\phi^+_d)$ and the bound \eqref{EAMentcriterion} which enables better noise-robustness.
For the special case of $n=d^2$ and $k=1$ our witness recovers the result introduced in \cite{ChenLiFei15}.

In summary, both entanglement criteria require no assumption on the state, they apply to states of any local dimension, they detect both the Schmidt number and the entanglement fidelity and their resource cost (the number of bases $m$ and the number of projectors $n$) can be freely selected by the experimenter. At this point, the natural question regards the usefulness of the criteria, i.e.~how good are they at detecting various forms of high-dimensional entanglement as compared to their resource cost. Next, we investigate this for important noise models.

\textit{Entanglement with depolarising noise.---} Consider that the source produces the maximally entangled state but is subjected to noise of uniform spectral density. The resulting isotropic state is $\rho_v^{\text{iso}}=v \ketbra{\phi^+_d}+\frac{1-v}{d^2}\openone$, where $v\in[0,1]$ is the visibility. This state has entanglement fidelity $F(\rho_v^{\text{iso}})=v+\frac{1-v}{d^2}$ and Schmidt number at least $k+1$ if and only if the visibility exceeds the critical value $v_\text{opt}=\frac{kd-1}{d^2-1}$  \cite{Terhal2000}. 

Using the MUB-criterion in Result~\ref{result1} on $\rho_v^{\text{iso}}$, the critical visibility for detecting Schmidt number at least $k+1$ becomes
\begin{equation}\label{visi1}
v_\text{MUB}=\frac{d-k+m(k-1)}{m(d-1)}.
\end{equation}
As expected, using more MUBs reduces $v_\text{MUB}$ for any $k$. For fixed $k$ and large $d$, the visibility threshold tends to $v_\text{MUB}=\frac{1}{m}$. Importantly, if we use a complete set of MUBs ($m=d+1$), then the criterion becomes necessary and sufficient as $v_\text{MUB}=v_\text{opt}$.

The key question is how rapidly $v_\text{MUB}$ becomes a good approximation of $v_\text{opt}$ as we increase $m$. Following \cite{Morelli2022}, we quantify this accuracy through the ratio of the relative gap between the two visibilities, defined as $\Delta\equiv \frac{1-v_\text{MUB}}{1-v_\text{opt}}$. Thus, $\Delta\approx 1$ ($\Delta\approx 0$) indicates a good (bad) approximation of the ideal value. One finds that $\Delta= \frac{d+1}{d}\left(1-\frac{1}{m}\right)$. Interestingly this is independent of $k$. Typically, we want to use a small subset of the total number of MUBs, i.e.~$m\ll d$. In this limit, we have $\Delta\approx 1-\frac{1}{m}$, which quickly approaches unit. For instance, regardless of $d$ and $k$, five ($20$) MUBs are needed to achieve $\Delta=0.8$ ($\Delta=0.95$).

Interestingly, the criterion~\ref{result2} performs very similarly. The threshold for Schmidt number at least $k+1$ via EAMs is
\begin{equation}\label{visi2}
v_\text{EAM}=\frac{d-k}{n-1}+\frac{k-1}{d-1}.
\end{equation}
Indeed, using larger EAMs reduces $v_\text{EAM}$ and using a maximal EAM ($n=d^2$) gives the ideal value $v_\text{EAM}=v_\text{opt}$.  In fact, by choosing $n=md+1-m$, our two criteria become identical, i.e.~$v_\text{EAM}=v_\text{MUB}$. Notably, since $d$ is significantly larger than $m$, we roughly have $n\approx md$. Thus, the number of global product projections is roughly the same for both criteria. The accuracy of the EAM-criterion for the isotropic state is $\Delta=\frac{(d+1)(n-d)}{d(n-1)}$, which has the same favourable scaling in $n$ as did the MUB-criterion.

\textit{Entanglement with dephasing noise.---} Let the source produce a maximally entangled state which is subjected dephasing noise in the computational basis, $\rho_v^\text{deph}=v \ketbra{\phi^+_d}+\frac{1-v}{d}\sum_{i=0}^{d-1} \ketbra{ii}{ii}$.
The Schmidt number is at least $k+1$ if and only if the visibility exceeds $v_\text{opt}=\frac{k-1}{d-1}$ \cite{Bavaresco2018}.

Since the noise appears in a specific basis, we must carefully choose specific MUBs when detecting the Schmidt number of $\rho_v^\text{deph}$ using Result~\ref{result1}. With a good choice, the minimal number of MUBs, namely $m=2$, suffices to obtain a necessary and sufficient criterion for all dimensions and all Schmidt numbers. Specifically, we choose the pair of MUBs consisting of the computational basis $\{\ket{a}\}_{a=0}^{d-1}$ and its Fourier transform $\{F\ket{a}\}_{a=0}^{d-1}$ where $F=\frac{1}{\sqrt{d}}\sum_{s,t=0}^{d-1}e^{\frac{2\pi i}{d}st}\ketbra{s}{t}$. When measuring the separable state $\varphi_d=\frac{1}{d}\sum_{i=0}^{d-1} \ketbra{ii}{ii}$, the first basis yields perfect correlations and the second basis yields uniformly random outcomes. Hence  $\mathcal{S}_{2,d}(\varphi)=1+\frac{1}{d}$. From Eq.~\eqref{MUBentcriterion}, the visibility threshold then becomes the solution to $2v+(1-v)\mathcal{S}_{2,d}(\varphi)=1+\frac{k}{d}$ which is identical to $v_\text{opt}$.

Let us now momentarily depart from the local filter picture and instead consider the case where all outcomes of a rank-1 measurement are resolved in each round. In that scenario, an interesting feature of Result~\ref{result2} is that it enables entanglement detection using the smallest number of outcomes. That is, we can detect entanglement from an EAM with just $n=d+1$. Naturally, it is expected that this  sparse approach gives a correspondingly poor tolerance to noise. However, it turns out to still be a useful approach for $\rho_v^\text{deph}$ in reasonably low dimensions.  We choose the EAM $\ket{\psi_a}=F\ket{\psi_a'}$ where  $\ket{\psi_a'}=\frac{1}{\sqrt{d}}\sum_{l=0}^{d-1} e^{\frac{2\pi i}{d+1}l(a-1)}\ket{l}$ for  $a=1,\ldots,d+1$. We then consistently find that $v_\text{EAM}=(d^2-3(d-k)-1)/(d^2-1)$. For instance, for $d=8$ we can detect entanglement when $v_\text{EAM}>\frac{2}{3}$ and even a maximal Schmidt number when $v\gtrsim 0.95$. Qualitatively similar results apply to the case of depolarising noise.

\textit{Entanglement with the worst-case noise.---} Consider a noisy state $\rho_v^{\text{worst}}=v \ketbra{\phi^+_d}+(1-v)\sigma$, where the state $\sigma$ is selected so that it has a maximal detrimental impact on entanglement detection for any given set of MUBs and EAMs. Thus, we must choose $\sigma$ so that it minimises $\mathcal{S}_{m,d}(\sigma)$ and $\mathcal{R}_{n,d}(\sigma)$ respectively. When maximal sets are used, namely $m=d+1$ and  $n=d^2$, the witness operators associated to \eqref{Wmub} and \eqref{Weam} both simplify to $\openone+d\ketbra{\phi^+_d}$, which implies $\mathcal{S}_{m,d}(\sigma)\geq 1$ and  $\mathcal{R}_{n,d}(\sigma)\geq 1$. However, when $m<d+1$ and $n<d^2$, the witness operators are not full rank and therefore we can always find a $\sigma$ such that $\mathcal{S}_{m,d}(\sigma)=0$ and  $\mathcal{R}_{n,d}(\sigma)=0$. With these worst-case choices, for maximal sets, we thus obtain the critical visibility  $v_\text{MUB}=v_\text{EAM}=\frac{k}{d}$ for Schmidt number $k+1$, which is identical to the exact fidelity criterion of Ref.~\cite{Terhal2000}. For non-maximal sets, we obtain 
\begin{align}
	&v_\text{MUB}=\frac{d+(m-1)k}{dm} \\
	& v_\text{EAM}=\frac{d-1}{n-1} +\frac{k(n-d)}{d(n-1)}.
\end{align}
As compared to the exact fidelity of the worst-case state, namely $F(\rho_v^{\text{worst}})=v$, our criteria exhibit similarly fast convergence properties as seen previously for isotropic noise. However, many entangled states fundamentally cannot be detected via fidelity estimation. An extreme example of such unfaithful entanglement amounts to choosing $\sigma=\ketbra{01}$. The state $\rho_v^{\text{worst}}$ is entangled for all $v>0$ but the quality of the exact fidelity bound is only $\Delta=1-\frac{1}{d}$ when $m=d+1$ or $n=d^2$. Naturally, the unfaithful property of this entanglement cannot be overcome by any fidelity estimation method.

\textit{Discussion.---}
We have developed criteria for detection and quantification of high-dimensional entanglement which combine several practically useful properties. We now discuss them one by one.

Firstly, the criteria require no assumption on the state. This comes with the advantage that they can be reliably applied without requiring that the experiment accurately follows a particular noise model, which is often difficult to determine anyway. 

Secondly, the criteria permit the experimenter to select the number of measurements used for entanglement detection.  Notably, when the state is expected to feature relatively little noise, a small number of measurement suffices to detect  high Schmidt numbers. To exemplify this, we have examined the data reported in \cite{HerreraValencia2020} from measuring two global product MUBs on a $19$-dimensional state. With our criterion then implies an entanglement fidelity of at least $92.7\%$ and a Schmidt number of at least $18$.

Thirdly, the criteria require only $md$ (MUBs) and $n$ (EAMs) global product projections, and knowledge of the count rate so that relative frequencies can be estimated. For a fixed number of MUBs, this scales only linearly in the dimension, thus improving significantly on the $d(d+1)$ projections needed for exact fidelity estimation with known count rate. Furthermore, it also improves on the $2d^2$ projections required in the two-MUB fidelity-based method of \cite{Bavaresco2018} and the $d^2+(m-1)d$  projections for $m$ MUBs from \cite{HerreraValencia2020} where the quadratic term is not used for determining the count rate but implicitly featured in the witness construction. For example for $m=5$ MUBs in $d=100$, the method used in \cite{HerreraValencia2020} would amount to $10400$ projections, as compared to our $500$. As we have shown, EAMs offer similar scaling advantages. 

Fourthly, for relevant types of noise the criteria rapidly become converge to the theoretically optimal noise-rates. We again exemplify this based on the data of \cite{HerreraValencia2020} where a $97$-dimensional state was measured in two MUBs. The correlation probability in the two respective bases are $0.6942$ and $0.6899$. A direct use of our criterion certifies $k\geq 38$. Suppose now that $m$ MUB-diagonals would have been measured in the experiment, and for sake of argument each achieving a correlation of $\frac{0.6942+0.6899}{2}=0.6920$. Measuring already one more MUB-diagonal ($m=3$) would then have implied $k\geq 53$,  whereas measuring all $m=98$ MUBs would only further improve this to $k\geq 67$.

Entanglement witnesses have become a standard tool in more and more complex experimental settings. While for each specific noise model and for each specific physical system, upon being able to find a good characterisation, special entanglement witnesses can be tailored, it is rare to find a universal criterion that retains a simplicitly in both its evaluation and its use, and more importantly in obtaining the relvant data. EAMs and especially MUBs are often the goal of many experiments. In some systems, MUBs are particularly natural choices, given the mutual unbiasedness of for example position and momemtum observables. The fact that high-dimensional entanglement can be detected and quantified in a noise-tolerant way using such commonplace measurements hopefully makes it a versatile tool to be used across many platforms.  

Finally, we note that our method does not straightforwardly extend to multipartite systems. In multipartite systems there is no counterpart to the Schmidt number and a unique maximally entangled state, as each entanglement class has a distinct maximally entangled state, so one must instead consider Schmidt numbers across given cuts of the system and orient the criterion towards a selected target state. More crucially, our bipartite criteria are valid regardless of which MUBs and EAMs are considered, and they are based on the idea of perfect correlations between two parties. The former's level of generality and the intuition behind the latter  idea do not carry over to multipartite systems. Multipartite entanglement criteria, in the spirit of our results, require more substantial innovation and therefore constitute a central open problem.

\begin{acknowledgements}
S.M. is supported by the Basque Government through IKUR strategy and through the BERC 2022-2025 program and by the Ministry of Science and Innovation: BCAM Severo Ochoa accreditation CEX2021-001142-S / MICIN / AEI / 10.13039/501100011033, PID2020-112948GB-I00 funded by MCIN/AEI/10.13039/501100011033 and by "ERDF A way of making Europe". M.H. acknowledges funding from the European Commission (grant ’Hyperspace’ 101070168) and from the European Research Council (Consolidator grant ’Cocoquest’ 101043705). A.T. is supported by the Wenner-Gren Foundation and  by the Knut and Alice Wallenberg Foundation through the Wallenberg Center for Quantum Technology (WACQT).
\end{acknowledgements}

\bibliography{SN_mubsic_references}

\appendix

\section{Proof of Result~1}
Define the witness operator $W=\sum_{z=1}^m A_z$ where $A_z=\sum_{a=0}^{d-1} \ketbra{e_a^z}{e_a^z}\otimes \ketbra{e_a^{z*}}{e_a^{z*}}$. Thus $\mathcal{S}_{m,d}=\Tr\left(W\rho_{AB}\right)$. For any linear operator $O$, it holds that $\openone \otimes O\ket{\phi^+_d}= O^T \otimes \openone \ket{\phi^+_d}$. Use this relation and the standard properties of a basis measurement  to obtain $A_z \ket{\phi^+_d}=\sum_a \ketbra{e_a^z}{e_a^z}\left( \ketbra{e_a^{z*}}{e_a^{z*}}\right)^\text{T} \otimes \openone\ket{\phi^+_d} =\sum_a\ketbra{e_a^{z}}{e_a^{z}}\otimes \openone \ket{\phi^+_d}= \ket{\phi^+_d}$. Hence, $\ket{\phi^+_d}$ is an eigenstate of $W$ with eigenvalue $m$,
\begin{equation}\label{mubeig}
W\ket{\phi^+_d}=m\ket{\phi^+_d}.
\end{equation}
Since $W\succeq 0$, it admits a spectral decomposition $W=\sum_{i=1}^{d^2} \lambda_i \ketbra{\lambda_i}{\lambda_i}$ where the set $\{\ket{\lambda_1},\ldots,\ket{\lambda_{d^2}}\}$ is the eigenbasis of $W$ and $\lambda_i\geq0$  is the eigenvalue associated to $\ket{\lambda_i}$. We can w.l.o.g.~select $\ket{\lambda_1}=\ket{\phi^+_d}$ and $\lambda_1=m$. Assume now that for all other eigenvalues it holds that $\lambda_i\leq 1$. Under this assumption, it follows from the completeness of the eigenbasis that
\begin{align}\nonumber\label{stepmub}
W=&(m-1)\phi^+_d +\left(\phi^+_d +\sum_{i=2}^{d^2} \lambda_i \ketbra{\lambda_i}{\lambda_i}\right) \\
&\leq (m-1)\phi^+_d +\openone.
\end{align}

Since $\mathcal{S}_{m,d}$ is a convex function in $\rho_{AB}$, we need only to consider pure states of Schmidt rank $k$ in order to prove a bound that holds for all states of Schmidt number $k$. Therefore let $\ket{\psi}$ be any state of Schmidt rank $k$. From \eqref{stepmub} we obtain
\begin{equation}\label{step}
\bracket{\psi}{W}{\psi}\leq 1+ (m-1) \left|\braket{\psi}{\phi^+_d}\right|^2\leq 1+\frac{(m-1)k}{d},
\end{equation}
where in the last step we have used the fidelity bound $F\leq \frac{k}{d}$. By noticing that $F(\psi)\geq \left|\braket{\psi}{\phi^+_d}\right|^2$, we can re-arrange the first inequality in \eqref{step} to obtain the bound on the entanglement fidelity.

To complete the proof we must show that the assumption $\lambda_i \leq 1$ holds. For this purpose, define the shifted witness operator $\tilde{W}=W-(m-1)\phi^+_d$. Its spectra reads $(1,\lambda_2,\ldots,\lambda_{d^2})$. We will show that $\tilde{W}$ is a projector ($\tilde{W}^2=\tilde{W}$), from which it immediately follows that all its eigenvalues belong to $\{0,1\}$. To prove that it is a projector, we compute its square,
\begin{equation}\label{squaremub}
\tilde{W}^2=W^2+(m-1)^2 \phi^+_d-(m-1)\{W,\phi^+_d\}.
\end{equation}
From Eq.~\eqref{mubeig}, it follows that $\{W,\phi^+_d\}=2m \phi^+_d$. Evaluating $W^2$ and using the MUB property one arrives at $W^2=W+\sum_{z\neq z'} \ketbra{\varphi_z}{\varphi_{z'}}$ where $\ket{\varphi_z}=\frac{1}{\sqrt{d}}\sum_{a=0}^{d-1} \ket{e_a^z,e_a^{z*}}$ is the maximally entangled state expressed in the $z$'th MUB. Since $\ket{\phi^+_d}$ is invariant under any unitary $U\otimes U^*$, it follows that all $\ket{\varphi_z}$ are identical and equal to $\ket{\phi^+_d}$. Hence, we arrive at $W^2=W+\sum_{z\neq z'}\phi^+_d=W+m(m-1)\phi^+_d$. Inserting these findings in \eqref{squaremub}, we arrive at
\begin{align}\nonumber
&\tilde{W}^2=W+m(m-1)\phi^+_d+(m-1)^2\phi^+_d\\\nonumber
&-2m(m-1)\phi^+_d=W-(m-1)\phi^+_d=\tilde{W},
\end{align}
which concludes the proof.

\section{Proof of Result~2}
The proof idea parallels that of Result~1. Due to the convexity of $\mathcal{R}_{n,d}$ in $\rho_{AB}$, we can restrict the analysis to pure states of Schmidt rank $k$.  We define the witness operator $W= \frac{d(n-1)}{n(d-1)} \sum_{a=1}^{n}\ketbra{\psi_a}\otimes \ketbra{\psi_a^*}$ and observe that 
\begin{equation}\label{siceig}
W\ket{\phi^+_d}=\frac{n-1}{d-1}\ket{\phi^+_d}.
\end{equation}
Hence the spectral decomposition becomes $W=\frac{n-1}{d-1}\phi^+_d+\sum_{i=2}^{d^2}\lambda_i \ketbra{\lambda_i}{\lambda_i}$, where $\{\lambda_i,\ket{\lambda_i}\}_{i=2}^{d^2}$ are the remaining eigenvalues and eigenvectors of $W$. Since $W\succeq 0$ we have $\lambda_{i}\geq 0$. If we assume that $\lambda_i\leq 1$, then
\begin{align}\nonumber
W&=\left(\frac{n-1}{d-1}-1\right)\phi^+_d+\left(\phi^+_d+\sum_{i=2}^{d^2}\lambda_i \ketbra{\lambda_i}{\lambda_i}\right)\\
&\leq \frac{n-d}{d-1}\phi^+_d+\openone.
\end{align}
Using this, for any pure state $\ket{\psi}$ of Schmidt rank $k$, we obtain 
\begin{equation}
\bracket{\psi}{R}{\psi}=1+\frac{n-d}{d-1}\left|\braket{\psi}{\phi^+_d}\right|^2\leq 1+k\frac{n-d}{d(d-1)},
\end{equation}
where in the last step we have used the fidelity bound $F\leq \frac{k}{d}$. Using that $F(\psi)\geq \left|\braket{\psi}{\phi^+_d}\right|^2$, re-arranging the first inequality gives the bound on the entanglement fidelity.

To complete the proof, we must prove that $\lambda_i\leq 1$. To do this, we define the shifted witness operator $\tilde{W}=W-\frac{n-d}{d-1}\phi^+_d$ and prove that it is a projector, thus implying that all its eigenvalues are confined to $\{0,1\}$. To this end, we evaluate the square of the operator, 
\begin{equation}\label{stepsic}
\tilde{W}^2=W^2+\left(\frac{n-d}{d-1}\right)^2\phi^+_d-\frac{n-d}{d-1}\{W,\phi^+_d\}.
\end{equation}
From Eq.~\eqref{siceig} it follows that the anticommutator reduces to $\frac{2(n-1)}{d-1}\phi^+_d$. Next we compute $W^2$,
\begin{align}\nonumber
W^2&=\left(\frac{d(n-1)}{n(d-1)}\right)^2 \sum_{a,a'} |\braket{\psi_a}{\psi_{a'}}|^2\ketbra{\psi_a}{\psi_{a'}}\otimes \ketbra{\psi_a^*}{\psi_{a'}^*}\\\nonumber
&=\frac{d(n-1)}{n(d-1)}W+\frac{d(n-d)(n-1)}{n^2(d-1)^2}\sum_{a\neq a'} \ketbra{\psi_a \psi_{a}^*}{\psi_{a'}\psi_{a'}^*}\\
&= W+\frac{d(n-d)(n-1)}{n^2(d-1)^2}\sum_{a} \ket{\psi_a \psi_a^*}\sum_{a'} \bra{\psi_{a'}\psi_{a'}^*}.
\end{align}
To arrive at the second line, we have used the defining equiangular relation. In the third line, the right-most operator is an unnormalised projector onto the bipartite superposition $\sum_{a} \ket{\psi_a \psi_a^*}$. In fact, its eigenstate is the maximally entangled state. To see that, we use the action-at-distance property of the maximally entangled state and the fact that the projectors in an EAM resolve the identity, 
\begin{align}\nonumber
&\sum_{a,a'} \ketbra{\psi_{a} \psi_a^*}{\psi_{a'}\psi_{a'}^*} \ket{\phi^+_d}=n\sum_{a}\ketbra{\psi_a}{\psi_a}\otimes \openone\ket{\phi^+_d}\\
&=n \times \frac{n}{d}\openone \otimes\openone\ket{\phi^+_d}=\frac{n^2}{d}\ket{\phi^+_d}.
\end{align}
We thus have $\sum_{a} \ket{\psi_a \psi_a^*}\sum_{a'} \bra{\psi_{a'}\psi_{a'}^*}=\frac{n^2}{d}\phi^+_d$. Inserting the above back into \eqref{stepsic}, we arrive at 
\begin{align}\nonumber
\tilde{W}^2&=W+\frac{d(n-d)(n-1)}{n^2(d-1)^2}\frac{n^2}{d}\phi^+_d+\frac{(n-d)^2}{(d-1)^2}\phi^+_d\\
&-\frac{2d(n-d)(n-1)}{(d-1)^2}\phi^+_d=W-\frac{n-d}{d-1}\phi^+_d=\tilde{W},
\end{align}
which concludes the proof.

\end{document}